\documentclass[floatfix,aps,prl,preprint,superscriptaddress]{revtex4-1}
\usepackage{graphicx}
\usepackage[]{epsfig}
\usepackage{times,amsmath,amssymb}

\usepackage{color}

\begin{document}

\title{Formation of quantum dots in GaN/AlGaN FETs}

\author{Tomohiro Otsuka}
\email[]{tomohiro.otsuka@riec.tohoku.ac.jp}
\affiliation{Research Institute of Electrical Communication, Tohoku University, 2-1-1 Katahira, Aoba-ku, Sendai 980-8577, Japan}
\affiliation{Center for Spintronics Research Network, Tohoku University, 2-1-1 Katahira, Aoba-ku, Sendai 980-8577, Japan}
\affiliation{Center for Science and Innovation in Spintronics, Tohoku University, 2-1-1 Katahira, Aoba-ku, Sendai 980-8577, Japan}
\affiliation{Center for Emergent Matter Science, RIKEN, 2-1 Hirosawa, Wako, Saitama 351-0198, Japan}

\author{Takaya Abe}
\affiliation{Research Institute of Electrical Communication, Tohoku University, 2-1-1 Katahira, Aoba-ku, Sendai 980-8577, Japan}

\author{Takahito Kitada}
\affiliation{Research Institute of Electrical Communication, Tohoku University, 2-1-1 Katahira, Aoba-ku, Sendai 980-8577, Japan}

\author{Norikazu Ito}
\affiliation{ROHM Co., Ltd, 21 Saiinnmizosakicho, Ukyo-ku, Kyoto, Kyoto 615-8585, Japan}

\author{Taketoshi Tanaka}
\affiliation{ROHM Co., Ltd, 21 Saiinnmizosakicho, Ukyo-ku, Kyoto, Kyoto 615-8585, Japan}

\author{Ken Nakahara}
\affiliation{ROHM Co., Ltd, 21 Saiinnmizosakicho, Ukyo-ku, Kyoto, Kyoto 615-8585, Japan}

\date{\today}

\begin{abstract}
GaN and the heterostructures are attractive in condensed matter science and applications for electronic devices.
We measure the electron transport in GaN/AlGaN field-effect transistors (FETs) at cryogenic temperature.
We observe formation of quantum dots in the conduction channel near the depletion of the 2-dimensional electron gas (2DEG).
Multiple quantum dots are formed in the disordered potential induced by impurities in the FET conduction channel.
We also measure the gate insulator dependence of the transport properties.
These results can be utilized for the development of quantum dot devices utilizing GaN/AlGaN heterostructures and evaluation of the impurities in GaN/AlGaN FET channels.
\end{abstract}

\maketitle

%\section{Introduction}

%GaN and GaN/AlGaN
GaN and the heterostructures are attractive materials because of their interesting electronic properties: the large direct bandgap, the high electron densities and mobilities.
They are utilized in light-emitting diodes~\cite{Akasaki1994, Nakamura1997, Akasaki2015}, power and high-frequency electronics devices~\cite{Mishra2002, Ikeda2010, Baliga2013}.
In electronic device applications, GaN/AlGaN heterostructures are important structures.
High density and high mobility 2DEG is formed at the interface~\cite{Ambacher1999, Manfra2004}.
The 2DEG is also investigated on the viewpoint of spin-orbit interactions~\cite{Thillosen2006, Schmult2006, Kurdak2006} and electron spin resonances~\cite{Shchepetilnikov2018}.
Quantum nanostructures can be fabricated from the heterostructure by utilizing nano-fabrications.
Quantum point contacts~\cite{Chou2005} and single electron transistors~\cite{Chou2006} are reported.
GaN/AlGaN nanowires~\cite{Risti2005, Songmuang2010} and self-assembled GaN islands~\cite{Nakaoka2007} are also used to form quantum dots.
Then GaN and the heterostructures are attractive also in quantum devices utilizing the electronic properties.

%Dopant quantum dots
Quantum dots can be formed also by intrinsic impurity potentials not only by the electric gates or edges defined structures.
In Si FETs, the formation of quantum dots by electrical potentials induced by dopants is reported~\cite{Sellier2006, Ono2007, Tabe2010}.
Dopants themselves work as quantum dots and control of the dopants~\cite{Tan2010} is used for quantum bit applications~\cite{Loss1997, Koppens2006, Yoneda2014, Veldhorst2015, Yoneda2018}, which is studied for quantum information processing~\cite{Nielsen2000, Ladd2010}. 
The stronger confinement by the dopant makes larger quantization energies and this enables high-temperature operation of the semiconductor quantum bits~\cite{Ono2019}.

%In this paper
In this paper, we measure electron transport through GaN/AlGaN FETs at cryogenic temperature.
We observe non-monotonic modulation of the current indicating formation of quantum dots near the pinch-off condition of the FET channel.
Multiple quantum dots are formed in the potential fluctuations induced by the impurities near the conduction channel.
We also measure the gate insulator dependence.

\section{Results}
\subsection{Device and FET properties}

\begin{figure}
\begin{center}
  \includegraphics{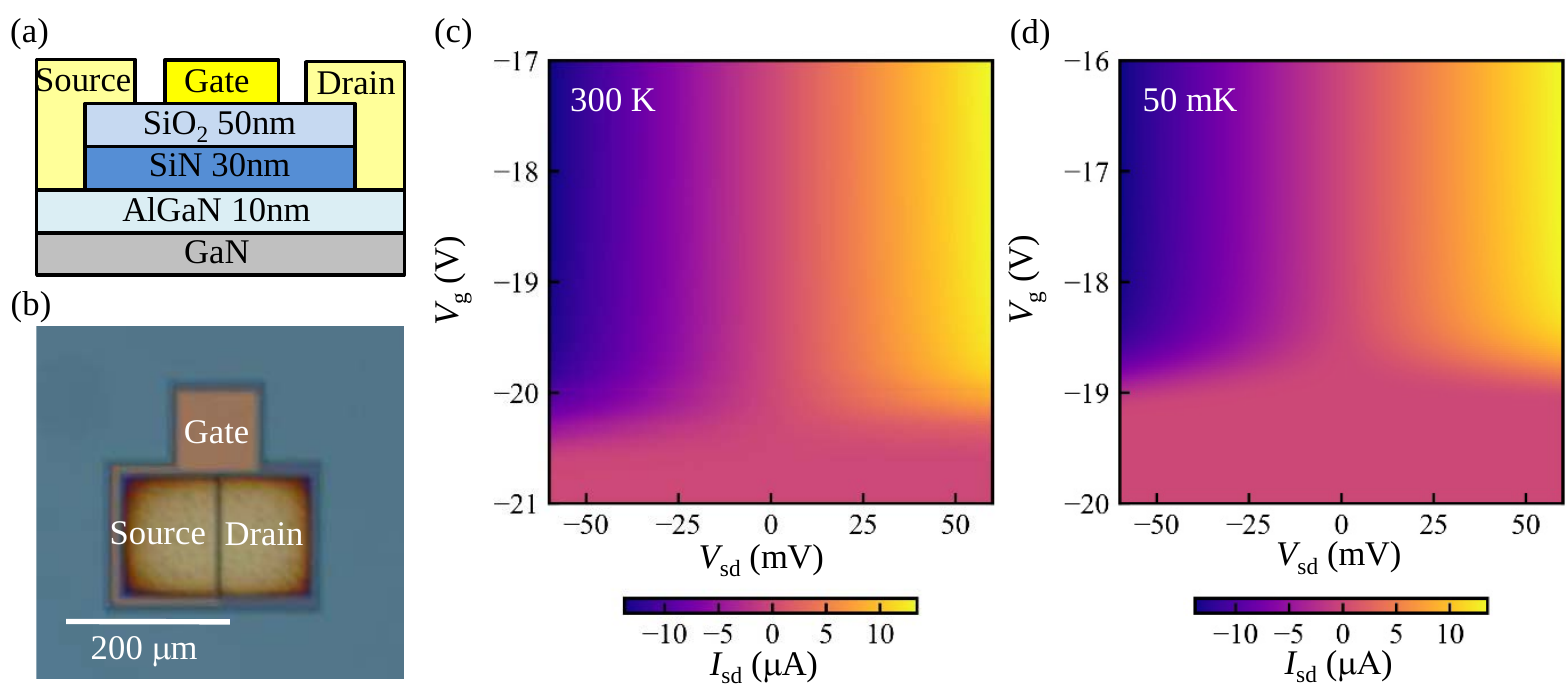}
  \caption{(a) Schematic of the layer structure of the device.
  2 DEG is formed at the interface between GaN and AlGaN.
  (b) Optical image of the device.
  The gate electrode with 1.4~$\mu $m gate length is placed between the source and the drain.
  (c), (d) Current trough the GaN/AlGaN FET as a function of the source-drain bias voltage $V_{\rm sd}$ and the gate voltage $V_{\rm g}$ at 300~K (c) and 50~mK (d).
  }
  \label{FET}
\end{center}
\end{figure}

%Device structure
Figure~\ref{FET}(a) shows a schematic of the layer structure of the device.
GaN and AlGaN layer is grown on the Si substrate by chemical vapor deposition.
At the interface between the GaN and AlGaN layers, 2DEG is formed.
The typical values of the electron density and the mobility are 6.7$\times 10^{12}$~cm$^{-2}$ and 1700~cm$^2$V$^{-1}$s$^{-1}$.
Source and drain contacts are prepared by Ti/Al.
A TiN gate electrode is deposited on the insulator of SiN and SiO$_2$.
SiN is grown in-situ just after the growth of the GaN/AlGaN.
An optical image of the device is Fig.~\ref{FET}(b).
The gate electrode is placed between the source and the drain contacts.
The gate length and the gate width are 1.4~$\mu $m and 150~$\mu $m, respectively.

The current between the source and the drain contacts $I_{\rm sd}$ is measured as a function of the applied source-drain bias voltage $V_{\rm sd}$ and the gate voltage $V_{\rm g}$.
We measure the current through the device at the room temperature 300~K and cryogenic temperature 50~mK.
The device is cooled down by a dilution refrigerator.

%FET properties
Figure~\ref{FET}(c) shows the measured current through the GaN/AlGaN FET $I_{\rm sd}$ at the room temperature 300~K.
In $V_{\rm g} > -20.5$~V, the FET channel is opened and the current flows depending on $V_{\rm sd}$. 
In the measurement, two 1~kOhm resistors, which is used for low pass filters designed for the cryogenic measurement, are inserted in series to the device and this limits the current in the open condition of the FET.
Around $V_{\rm g} \sim -20.5$~V, the conduction channel is depleted.
No current flows in more negatively gated region $V_{\rm g} < -20.5$~V.

Figure~\ref{FET}(d) shows the measured $I_{\rm sd}$ at the cryogenic temperature 50~mK.
The conduction channel remains at this temperature in $V_{\rm g} > -19$~V.
The depletion of the conduction channel occurs around $V_{\rm g} \sim -19$~V.
The pinch-off voltage shifts 1.5~V positively compared to the result at the room temperature.
This is induced by the suppression of the thermally induced carriers at the cryogenic temperature.

\subsection{Formation of quantum dots}

\begin{figure}
\begin{center}
  \includegraphics{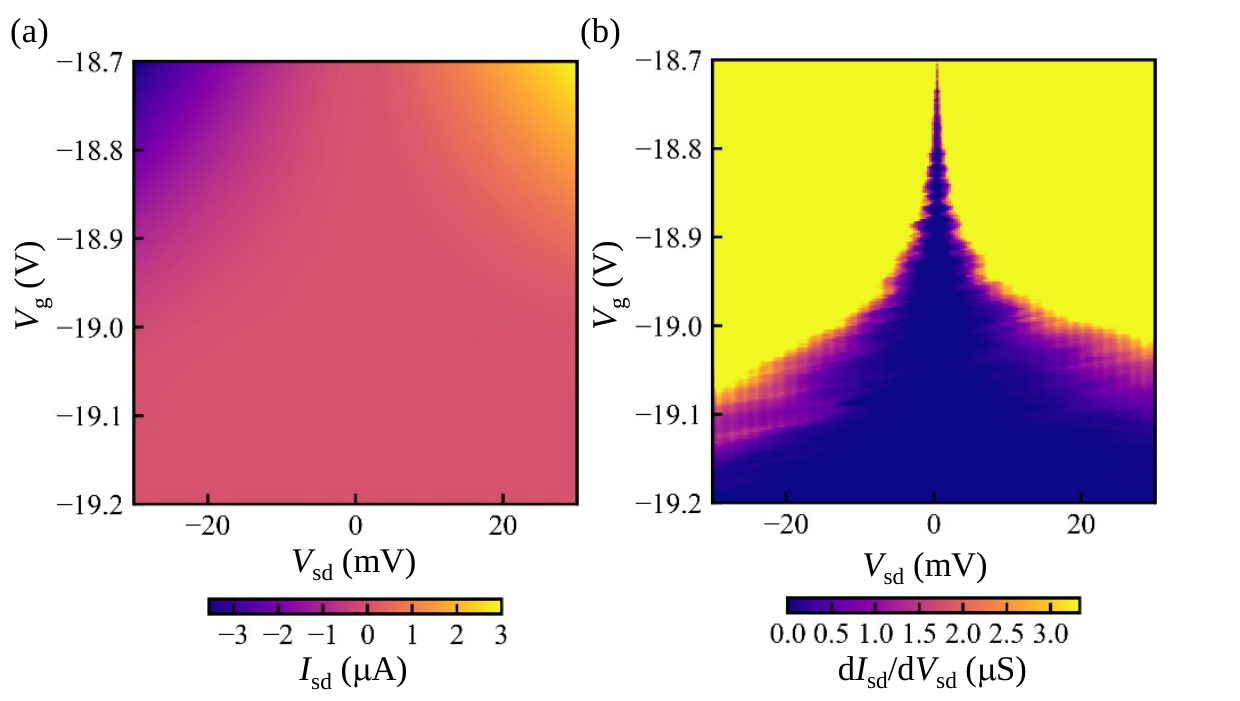}
  \caption{(a) Current through the FET as a function of the source-drain bias voltage $V_{\rm sd}$ and the gate voltage $V_{\rm g}$ at 50~mK near the depletion condition of the 2DEG.
  (b) The numerical derivative of the measured current as a function of the source-drain bias voltage ${\rm d}I_{\rm sd}/{\rm d}V_{\rm sd}$. 
  Non-monotonic modulation of the current and Coulomb diamond structures are observed.
  }
  \label{PinchOff}
\end{center}
\end{figure}

Figure~\ref{PinchOff}(a) shows the current through the FET near the depletion condition of the 2DEG.
The current is suppressed around the zero bias and non-linear I-V properties are observed in this region.
A numerical derivative of the measured current as a function of the source-drain bias voltage ${\rm d}I_{\rm sd}/{\rm d}V_{\rm sd}$ is shown in Fig.~\ref{PinchOff}(b).
The current $I_{\rm sd}$ is blocked around the zero bias condition $V_{\rm sd}\sim 0$.
The width of the blocked region is modulated by the gate voltage $V_{\rm g}$ and Coulomb diamonds are observed.
The size of the diamonds becomes larger in more negative values of $V_{\rm g}$ and this reflects that the dot size becomes smaller and the charging energy becomes larger.
Note that the faint vertical lines around the outside of the diamonds are the measurement artifact that originates from the output voltages of the source measure unit used in this measurement.

\begin{figure}
\begin{center}
  \includegraphics{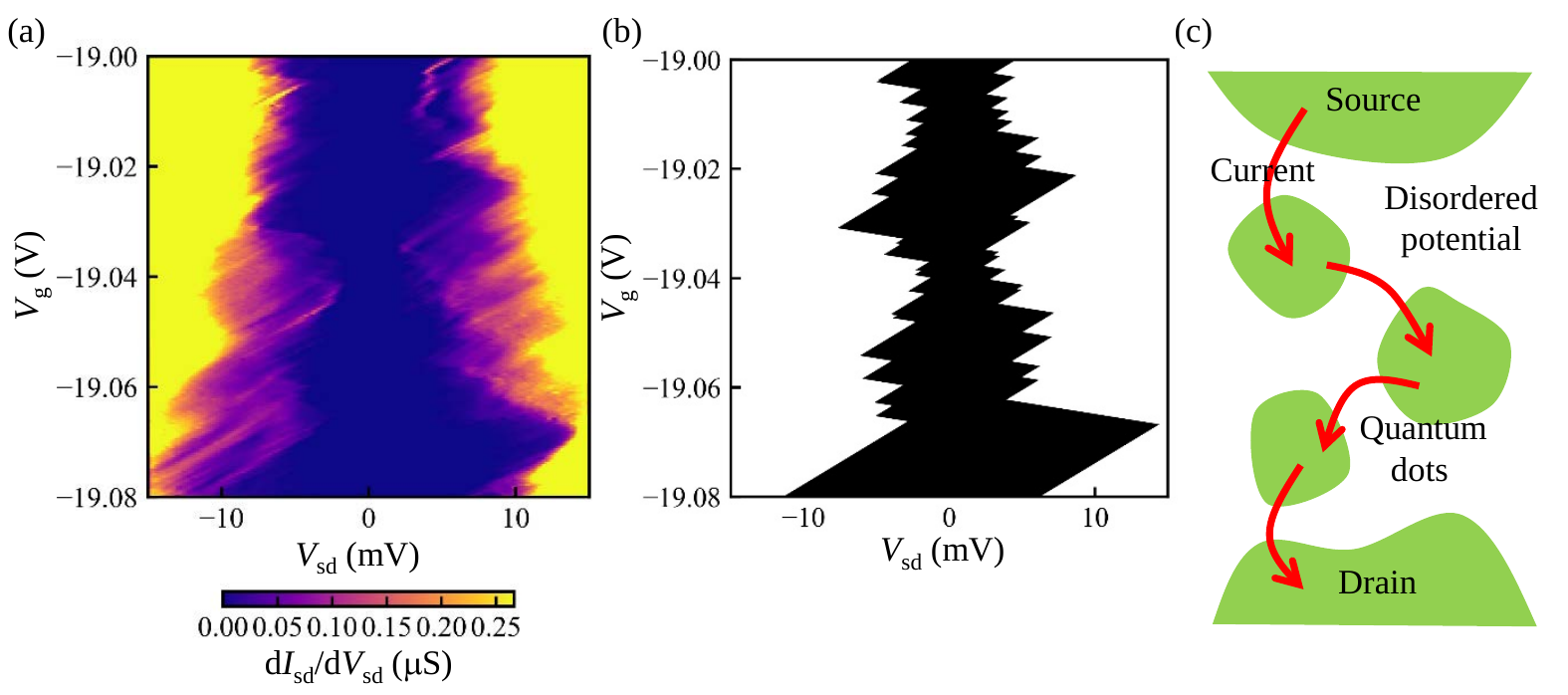}
  \caption{(a) Numerical derivative of the current as a function of the source-drain bias voltage.
  Coulomb diamonds are observed.
  (b), (c) Schematic of the one possible configuration of the quantum dots (c) and the expected Coulomb diamonds (b).
  Electrostatic potential will be disordered by the impurities and defects and the quantum dots are formed at the potential minima.
  Here we assumed that three quantum dots coupled in series and the overlapped Coulomb diamonds show gaps around the zero bias conditions.
  }
  \label{Fig3}
\end{center}
\end{figure}

Figure~\ref{Fig3}(a) shows the closed up image of the Coulomb diamonds.
In this small current condition, we use a current preamplifier to measure the current instead of the source measure unit and the measurement artifact like in Fig.~\ref{PinchOff}(b) is not there.
The current enhancement by the excited states is also observed as lines outside of the Coulomb diamonds.
Quantum dots are formed in the conduction channel of the FET.

The visible lines mostly have the same slope and this indicates that the dot is asymmetrically coupled to the leads: the dot is strongly coupled to one of the leads.
The voltage drop by forming the large in-series resistance in the conduction channel is negligible, which can be evaluated by inverting the source and the drain contacts in the measurement~\cite{OnoAPL2013}. 
The diamonds are not completely closed at $V_{\rm sd} = 0$ in Fig.~\ref{Fig3}(a).
This shows that multiple quantum dots are formed in this device.

\section{Discussion}

There are no small fine gates or structures to define quantum dots intentionally in this device.
The quantum dots will be formed by the disordered potential induced by the impurities or defects near the conduction channel.
Near the depletion of the 2DEG, the potential minima of the disordered potential contribute to the transport and coupled quantum dots are formed.

Figure~\ref{Fig3}(c) is a schematic of one possible configuration of the formed quantum dots.
Three quantum dots are coupled in series.
The resulting Coulomb diamonds become the overlap of the diamonds of each dot in a simple approximation~\cite{Nuryadi2003}.
Figure~\ref{Fig3}(b) shows the result when we assume the three quantum dots with charging energies $E_{\rm C1} , E_{\rm C2}, E_{\rm C3} =2.6, 2.3, 3.0$~meV, orbital level spacing $\Delta \epsilon_{\rm 1} , \Delta \epsilon_{\rm 2}, \Delta \epsilon_{\rm 3} = 0.97-11, 0.81-4.1, 0.65-2.9$~meV and alfa factor $\alpha = 0.0079$.
The black area indicates the Coulomb blocked region.
The model capture the main feature of Fig.~\ref{Fig3}(a).

\begin{figure}
\begin{center}
  \includegraphics{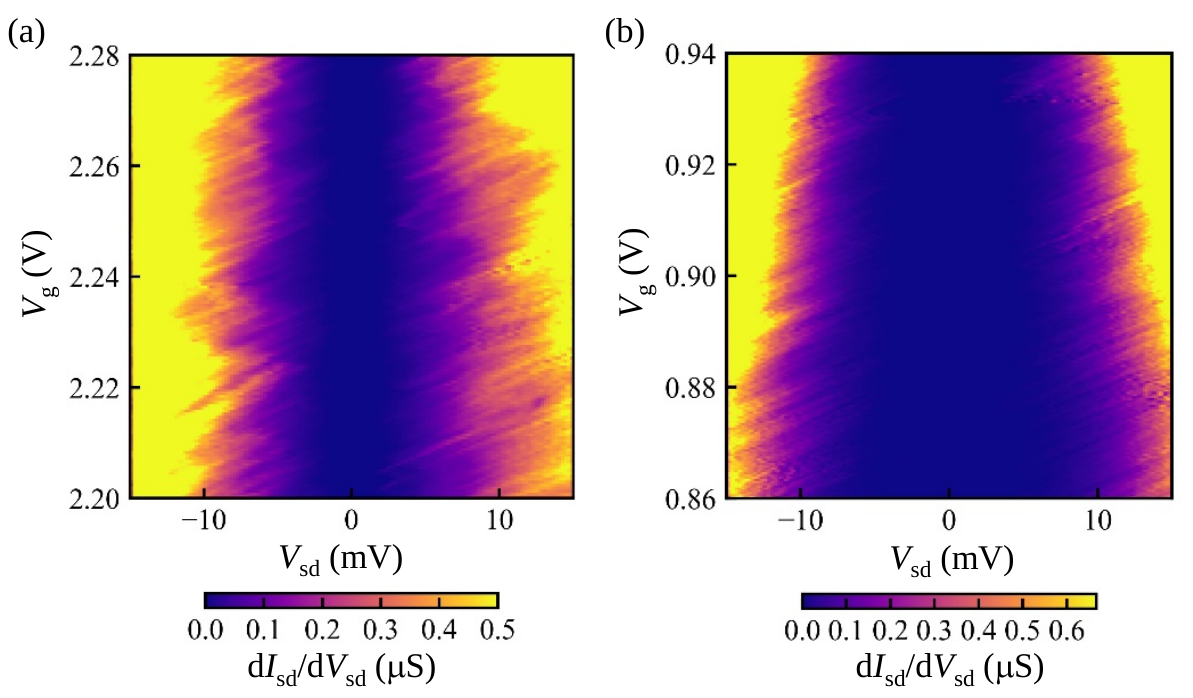}
  \caption{(a), (b) Numerical derivative of the current as a function of the source-drain bias voltage observed in other samples with different insulators SiO$_2$ (a) and SiN/SiO$_2$. 
  }
  \label{Fig4}
\end{center}
\end{figure}

To study the growth condition dependence of the quantum dot formation, we measure other samples with different gate insulators and fabrication processes, which induce different disorder densities.
In these new samples, the insulator is fabricated after taking out the samples from the growth chamber of GaN/AlGaN and etching processes.
The gate length is 0.6~$\mu $m.
Higher disroder densities are expected compared to the previous sample which has SiN insulators grown in-situ in the same chamber. 
Figure~\ref{Fig4}(a) and (b) show the results measured in devices with SiO$_2$ and SiN/SiO$_2$ insulators, respectively.
Compared to Fig.~\ref{Fig3}(a), more Coulomb diamonds are overlapped and the larger opening of the gap around the zero bias condition is observed.
More quantum dots are formed and coupled in series.
This is consistent with the expectation that the higher disorder density forms more quantum dots in these devices.
These support that the origin of the formation of the quantum dots is the disordered potentials around the FET channels.

In conclusion, we measure electron transport in GaN/AlGaN FETs at cryogenic temperature.
Quantum dots are formed in the conduction channel near the depletion of the 2DEG.
Multiple quantum dots are formed by the disordered potential in the FET.
We also measured insulator dependence of the quantum dot formation.
These results can be utilized for the development of quantum dot devices like semiconductor quantum bits and nano-probes~\cite{Altimiras2010, Otsuka2017, Otsuka2019} utilizing GaN/AlGaN and evaluation of the disordered potential in GaN/AlGaN FET channels.

\section{Acknowledgements}

We thank Takeshi Kumasaka for fruitful discussions and technical supports.
Part of this work is supported by
ROHM Collaboration Project,
PRESTO (JPMJPR16N3), JST, 
Futaba Electronics Memorial Foundation Research Grant, 
Iketani Science and Technology Foundation Research Grant,
Yamaguchi Foundation Research Grant,
The Mikiya Science and Technology Foundation Research Grant,
Harmonic Ito Foundation Research Grant,
Takahashi Industrial and Economic Research Foundation Research Grant,
The Murata Science Foundation Research Grant,
Samco Foundation Research Grant,
Casio Science Promotion Foundation Research Grant.

\section{Author contributions}
T. O. and K. N. planned the project; 
N. I., T. T., and K. N. performed device fabrication; 
T. O., T. A., T. K., N. I., T. T., and K. N. conducted experiments and data analysis; 
all authors discussed the results; 
T. O., T. A., T. K., T. T., and K. N. wrote the manuscript.

\section{Additional information}
Competing financial interests: The authors declare no competing financial interests.


\begin{references}

\bibitem{Akasaki1994}
Akasaki, I. \& Amano, H. Widegap Column-Ill Nitride Semiconductors for UV/Blue Light Emitting Devices. J. Electrochem. Soc. 141, 2266-2271 (1994).

\bibitem{Nakamura1997}
Nakamura, S. \& Fasol, G. The blue laser diode. (Springer, Berlin, 1997).

\bibitem{Akasaki2015}
Akasaki, I. Fascinating journeys into blue light (Nobel Lecture). Ann. Phys. 527, 311-326 (2015).

\bibitem{Mishra2002}
Mishra, U. K., Parikh, P. \& Wu, Y. F. AlGaN/GaN HEMTs - An overview of device operation and applications. Proc. IEEE 90, 1022-1031 (2002).

\bibitem{Ikeda2010}
Ikeda, N. et al. GaN power transistors on si substrates for switching applications. Proc. IEEE 98, 1151-1161 (2010).

\bibitem{Baliga2013}
Baliga, B. J. Gallium nitride devices for power electronic applications. Semicond. Sci. Technol. 28, 074011-1-8 (2013).

\bibitem{Ambacher1999}
Ambacher, O. et al. Two-dimensional electron gases induced by spontaneous and piezoelectric polarization charges in N- And Ga-face AIGaN/GaN heterostructures. J. Appl. Phys. 85, 3222-3233 (1999).

\bibitem{Manfra2004}
Manfra, M. J. et al. Electron mobility exceeding 160000 cm 2/V s in AlGaN/GaN heterostructures grown by molecular-beam epitaxy. Appl. Phys. Lett. 85, 5394-5396 (2004).

\bibitem{Thillosen2006}
Thillosen, N. et al. Weak antilocalization in gate-controlled Alx Ga1-x N GaN two-dimensional electron gases. Phys. Rev. B 73, 241311-1-4 (2006).

\bibitem{Schmult2006}
Schmult, S. et al. Large Bychkov-Rashba spin-orbit coupling in high-mobility GaN Alx Ga1-x N heterostructures. Phys. Rev. B 74, 033302-1-4 (2006).

\bibitem{Kurdak2006}
Kurdak, C., Biyikli, N., Ozgur, U., Morkoc, H. \& Litvinov, V. I. Weak antilocalization and zero-field electron spin splitting in Alx Ga1-x N AlN GaN heterostructures with a polarization-induced two-dimensional electron gas. Phys. Rev. B 74, 113308-1-4 (2006).

\bibitem{Shchepetilnikov2018}
Shchepetilnikov, A. V. et al. Electron spin resonance in a 2D system at a GaN/AlGaN heterojunction. Appl. Phys. Lett. 113, 052102-1-3 (2018).

\bibitem{Chou2005}
Chou, H. T. et al. High-quality quantum point contacts in GaNAlGaN heterostructures. Appl. Phys. Lett. 86, 073108-1-3 (2005).

\bibitem{Chou2006}
Chou, H. T. et al. Single-electron transistors in GaN/AlGaN heterostructures. Appl. Phys. Lett. 89, 033104-1-3 (2006).

\bibitem{Risti2005}
Risti\'{c}, J. et al. Columnar AlGaN/GaN nanocavities with AlN/GaN bragg reflectors grown by molecular beam epitaxy on Si(111). Phys. Rev. Lett. 94, 146102-1-4 (2005).

\bibitem{Songmuang2010}
Songmuang, R. et al. Quantum transport in GaN/AlN double-barrier heterostructure nanowires. Nano Lett. 10, 3545-3550 (2010).

\bibitem{Nakaoka2007}
Nakaoka, T., Kako, S., Arakawa, Y. \& Tarucha, S. Coulomb blockade in a self-assembled GaN quantum dot. Appl. Phys. Lett. 90, 162109-1-3 (2007).

\bibitem{Sellier2006}
Sellier, H. et al. Transport spectroscopy of a single dopant in a gated silicon nanowire. Phys. Rev. Lett. 97, 206805-1-4 (2006).

\bibitem{Ono2007}
Ono, Y. et al. Conductance modulation by individual acceptors in Si nanoscale field-effect transistors. Appl. Phys. Lett. 90, 102106-1-3 (2007).

\bibitem{Tabe2010}
Tabe, M. et al. Single-Electron Transport through Single Dopants in a Dopant-Rich Environment. Phys. Rev. Lett. 105, 016803-1-4 (2010).

\bibitem{Tan2010}
Tan, K. Y. et al. Transport Spectroscopy of single phosphorus donors in a silicon nanoscale transistor. Nano Lett. 10, 11-15 (2010).

\bibitem{Loss1997}
Loss, D., DiVincenzo, D. P. \& DiVincenzo, P. Quantum computation with quantum dots. Phys. Rev. A 57, 120-126 (1997).

\bibitem{Koppens2006}
Koppens, F. H. L. et al. Driven coherent oscillations of a single electron spin in a quantum dot. Nature 442, 766-771 (2006).

\bibitem{Yoneda2014}
Yoneda, J. et al. Fast electrical control of single electron spins in Quantum dots with vanishing influence from nuclear spins. Phys. Rev. Lett. 113, 267601-1-5 (2014).

\bibitem{Veldhorst2015}
Veldhorst, M. et al. A two-qubit logic gate in silicon. Nature 526, 410-414 (2015).

\bibitem{Yoneda2018}
Yoneda, J. et al. A quantum-dot spin qubit with coherence limited by charge noise and fidelity higher than 99.9\%. Nat. Nanotechnol. 13, 102-106 (2018).

\bibitem{Nielsen2000}
Nielsen, M. A. \& Chuang, I. L. Quantum Computation and Quantum Information. (Cambridge University Press, 2000). 7.

\bibitem{Ladd2010}
Ladd, T. D. et al. Quantum computers. Nature 464, 45-53 (2010).

\bibitem{Ono2019}
Ono, K., Mori, T. \& Moriyama, S. High-temperature operation of a silicon qubit. Sci. Rep. 9, 469-1-8 (2019).

\bibitem{OnoAPL2013}
Ono K., Tanamoto T. \& Ohguro T. Pseudosymmetric bias and correct estimation of Coulomb/confinement energy for unintentional quantum dot in channel of metal-oxide-semiconductor field-effect transistor. Appl. Phys. Lett. 103, 183107-1-4 (2013).

\bibitem{Nuryadi2003}
Nuryadi, R., Ikeda, H., Ishikswa, Y. \& Tabe, M. Ambipolar Coulomb blockade characteristics in a two-dimensional Si multidot device. IEEE Trans. Nanotechnol. 2, 231–235 (2003).

\bibitem{Altimiras2010}
Altimiras, C. et al. Tuning energy relaxation along quantum hall channels. Phys. Rev. Lett. 105, 226804-1-4 (2010).

\bibitem{Otsuka2017}
Otsuka, T. et al. Higher-order spin and charge dynamics in a quantum dot-lead hybrid system. Sci. Rep. 7, 12201-1-7 (2017).

\bibitem{Otsuka2019}
Otsuka, T. et al. Difference in charge and spin dynamics in a quantum dot-lead coupled system. Phys. Rev. B 99, 085402-1-5 (2019).

\end{references}
\end{document}